\documentclass[aps,twocolumn,prl,longbibliography]{revtex4-1}
\usepackage{graphicx,dcolumn,bm,float,longtable,mathptmx}
\usepackage{amsfonts,amsmath,amssymb}
\usepackage[usenames,dvipsnames]{xcolor}

\begin{document}


\title{Theory of perpendicular magnetocrystalline anisotropy in Fe/MgO (001)}

\author{Dorj Odkhuu$^{1,2}$}
\author{Won Seok Yun$^{1,3}$}
\author{S. H. Rhim$^{1}$}
\email[Email address: ]{sonny@ulsan.ac.kr}
\author{Soon-Cheol Hong$^{1}$}
\email[Email address: ]{schong@ulsan.ac.kr}
\affiliation{
  $^{1}$ Department of Physics and Energy Harvest Storage Research Center,
  University of Ulsan, Ulsan, 680-749, Republic of Korea\\
  $^{2}$ Department of Physics,
  Incheon National University, Incheon, 406-772, Republic of Korea\\
  $^{3}$ Department of Emerging Materials Science, DGIST, Daegu, 711-873, Republic of Korea
}

\date{\today}

\begin{abstract}
  The origin of large perpendicular magneto-crystalline anisotropy (PMCA)
  in Fe/MgO (001) is revealed by comparing Fe layers with and without the MgO.
  Although Fe-O $p$-$d$ hybridization is weakly present,
  it cannot be the main origin of the large PMCA as claimed in previous study.
  Instead, perfect epitaxy of Fe on the MgO is more important to achieve such large PMCA.
  As an evidence,
  we show that the surface layer in a clean free-standing Fe (001) dominantly contributes to $E_{MCA}$,
  while in the Fe/MgO, those by the surface and the interface Fe layers contribute almost equally.
  The presence of MgO does not change positive contribution from $\langle xz|\ell_Z|yz\rangle$,
  wherease it reduces negative contribution from $\langle z^2|\ell_X|yz\rangle$ and $\langle xy|\ell_X|xz,yz\rangle$.
\end{abstract}

\maketitle


Exploration for magnetic materials with future applications dates back
more than two decades,
which includes giant magnetoresistance (GMR) and many applications in spintronics
such as magneto-resistive random-access-memory (MRAM), magnetic sensors,
and novel programmable logic devices\cite{wolf:sci01}.
Among those, perpendicular magnetocrystalline anisotropy (PMCA)
has attracted greatly as materials with large PMCA can offer 
more opportunities to realize magnetic devices. 
It can provide an ideal tool to realize spin transfer torque excluding external fields.
Also it offers large bit density in practical applications.
Magnetic tunnel junctions (MTJs) using MgO as an electrode, in particular,
have drawn huge attention
owing to enormous magnetoresistance reaching as high as 180\%\cite{yuasa04:nmat}.
Furthermore, large PMCA in Fe/MgO has shown great promise
with tremendous performance\cite{ikeda10:nmat,brataas12:nmat,mangin06:nmat}.

There have been numerous subsequent works ever since including Fe/MgO and FeCoB$\mid$MgO.
In theoretical sides, several ways have been proposed to enhance PMCA in Fe/MgO:
By applying an external electric field\cite{kohji09:PRL,kohji10:prb},
adding heavy transition metal layer on top of Fe\cite{odkhuu13:prb}, and so forth.
While it is still on-going endeavor to enhance PMCA,
it looks quite indispensable to identify the physics origin of PMCA in Fe/MgO without aforementioned effects.
Although it has been attributed to the $p$-$d$ hybridization
between Fe and O atoms at the interface\cite{yang11:prb,hallal13:prb} that is responsible for large PMCA,
we argue that the hybridization
cannot be the main driving force.
If the hybridization is indeed the main source of PMCA,
why PMCA is absent in Fe grown on AlO$_x$\cite{miyazaki95:jmmm,wang04:IEEE} 
despite the presence of apparent hybridization between Fe and O?
Furthermore, if the hybridization is really the main origin of PMCA,
then it should be small when Fe/MgO is underoxidized
due to the reduction of hybridization\cite{yang11:prb}.

In this paper, we show that the most dominant contribution to PMCA is not the $p$-$d$ hybridization
despite its weak presence in Fe/MgO.
Instead, perfect epitaxy
of the interface would be the key, which is an indicator of high level of sample fabrication.
To support our idea, the contribution from the surface
and the interface Fe in the presence of the MgO
is compared, which is almost the same in magnitude.
This clearly counter-argues that the hybridization plays the main role of PMCA,
whose detailed analysis of the electronic structure is provided.

\begin{table*}[t]
  \centering
  \caption{Magnetic moments (in $\mu_{B}$) of Fe atoms in the clean Fe and Fe/MgO,
    where Fe(S), Fe(S-1), Fe(C), and Fe(I) stand for Fe atom
    at surface, subsurface, center, and interface, respectively.
    Without MgO, Fe(S-1) and Fe(C) are not well defined for 1-, 2-ML, so is Fe(S-1) for 3-ML.}
  \label{tab:1}
  \begin{ruledtabular}
  \begin{tabular}{{c|ccc|cccc}}
\# of Fe layers &  \multicolumn{3}{c|}{clean Fe}  & \multicolumn{4}{c}{Fe/MgO}\\
\hline
       & Fe(S) & Fe(S-1) & Fe(C) & Fe(S) & Fe(S-1) & Fe(I) & O(I)\\
       \hline
    1 & 3.14 & --   &  --  & --   &  --  & 3.09 & 0.03 \\
    2 & 2.82 & --   &  --  & 2.85 &  --  & 2.62 & 0.02 \\
    3 & 2.96 & 2.39 &  --  & 2.96 & 2.45 & 2.83 & 0.02 \\
    4 & 2.94 & 2.45 &  --  & 2.92 & 2.44 & 2.72 & 0.02 \\
    5 & 2.96 & 2.43 & 2.62 & 2.95 & 2.44 & 2.80 & 0.02 \\
    6 & 2.95 & 2.44 & 2.56 & 2.95 & 2.44 & 2.79 & 0.02 \\
    7 & 2.95 & 2.43 & 2.42 & 2.95 & 2.44 & 2.79 & 0.02 \\
  \end{tabular}
\end{ruledtabular}
\end{table*}

\begin{figure}[b]
  \centering
  \includegraphics[width=\columnwidth]{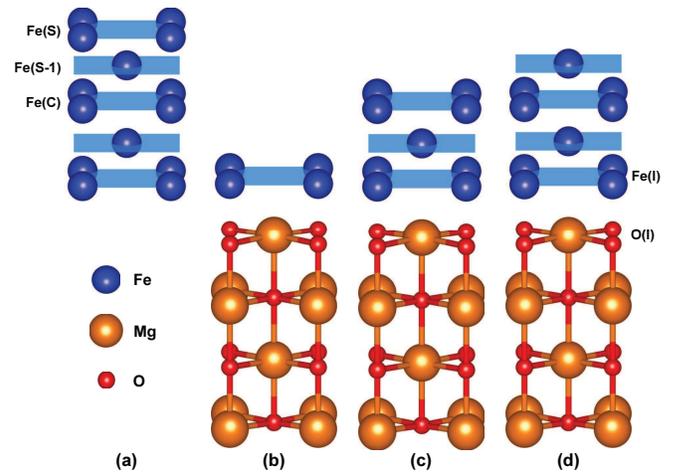}
\caption{(color online) (a) MgO-free Fe layers  and Fe/MgO with substrate with (b) 1-ML, 
(c) 3-ML, and (d) 4-ML Fe. 
Spheres with yellow, red, and green color denote Fe, Mg, and O atoms, respectively.
S and I denote atoms at the surface and the interface, S-1 that of subsurface.}
  \label{fig:1}
\end{figure}
Density functional calculations are performed using 
the highly precise full-potential linearized augmented plane wave (FLAPW) method\cite{wimmer:81}
and Vienna {\em Ab-initio} Simulation Package (VASP)\cite{kresse93:vasp,kresse96:vasp}.
Generalized gradient approximation (GGA)
by Perdew, Burke, and Ernzerhof (PBE) parametrization\cite{perdew96:prl}
is employed for the exchange-correlation potential.
In VASP calculations with projector augmented basis sets\cite{bloechl94:PAW},
full atomic relaxations have been carried out within force criteria 0.001 eV/\AA,
where cutoff energy 400 eV for wave function expansion
and 16$\times$16$\times$1 {\em k} mesh in the irreducible Brillouin zone wedge are chosen.
In FLAPW calculations,
cutoffs for wave function and potential representations are
16 and 256 Ry, respectively.
Charge densities and potential inside muffin-tin (MT) spheres
were expanded with lattice harmonics $\ell \leq 8 $
with MT radii of 2.1, 1.4, and 1.8 a.u. for Fe, O, and Mg atoms, respectively.
For {\em k} point summation,
we used 24$\times$24$\times$1 mesh in the irreducible Brillouin zone wedge. 
A self-œôø²consistent criteria of 1.0$\times$10$^{œôø²-5}$ $e$/(a.u.)$^3$ was imposed,
where convergence with respect to the numbers of the basis functions
and {\em k} points was also seriously checked.
MCA energies ($E_{MCA}$) are obtained using FLAPW.
For the calculation of $E_{MCA}$,
torque method\cite{wang96:prb_torque} was employed to reduce computational costs,
whose validity and accuracy have been successfully proved
in conventional FM materials and others\cite{wang93:prb,wu99:jmmm,zhang10:prb82,odkhuu11:apl,odkhuu13:prb,odkhuu15:prb,hotta13:prl,hotta13:jkps}

The model geometry in our study is depicted in Fig.~\ref{fig:1}: 
(a) MgO-free Fe-layers and (b-d) Fe-layers on the MgO substrate,
where we have considered number of Fe layers from 1 to 7 and five layers of MgO.
The experimental lattice constant of MgO (4.214~\AA) was adopted for the in-plane lattices.
Magnetic moments of Fe atoms are listed in Table~\ref{tab:1},
where Fe atoms at the surface,  subsurface, (one layer beneath the topmost surface),
and the interface with MgO, are denoted by Fe(S), Fe(S-1), and Fe(I),
respectively, as in Fig.~\ref{fig:1}.
The center layer in the absence of MgO is labelled as Fe(C).
Since Fe(S) and Fe(I) are identical 1-ML Fe/MgO, only Fe(I) is shown.
In the clean Fe layers, i.e. MgO-free Fe layers,
moments of Fe(S) are enhanced with respect to the bulk
giving maximum value of 3.14 $\mu_{B}$ for 1-ML Fe.
Moments of Fe(S-1) are comparable or smaller than those of Fe(C).
Fe(C) moments are slightly larger than bulk Fe, which is due to enlarged lattice constants.
The presence of MgO also enhances moments of Fe(S) up to 2.96 $\mu_{B}$.
In all cases, Fe(S) have the largest moments mostly around 2.95 $\mu_{B}$.
Interestingly, Fe(I) moments are
smaller than but comparable to those of Fe(S).
This feature has been already addressed in previous work\cite{li91:prb}.

\begin{figure}[b]
  \centering
  \includegraphics[width=\columnwidth]{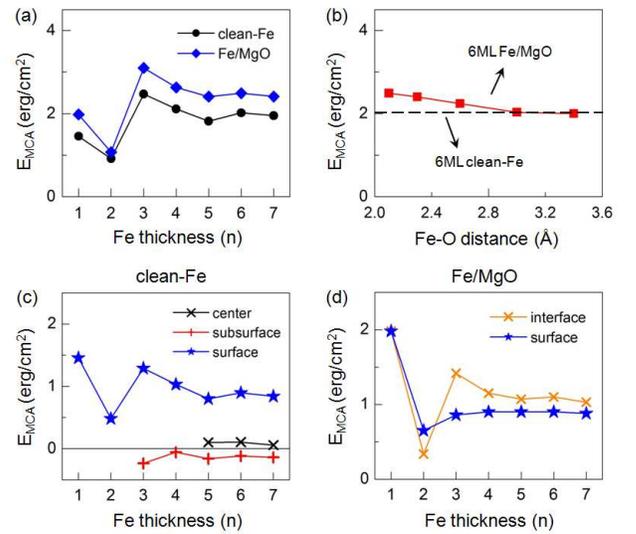}
  \caption{(a) $E_{MCA}$ as function of number of Fe layers
    with (blue line) and without MgO (black line).
    (b) $E_{MCA}$ as a function of Fe-O distance for 6-Fe/MgO.
    Atomic resolution of $E_{MCA}$ versus number of Fe layer for the (c) clean Fe and (d) Fe/MgO
}
  \label{fig:2}
\end{figure}
$E_{MCA}$ as function of the number of Fe layers is plotted in Fig.~\ref{fig:2}
with and without MgO.
The presence of MgO systematically increases total $E_{MCA}$ by $\sim$20\%
with respect to the free-standing Fe-layer except 2-ML Fe.
As seen clearly, the saturation behavior is evident as the number of Fe layer increases.
The singular behavior of 2-ML Fe will be discussed later.
To reveal the role of Fe 3$d$-O 2$p$ hybridization,
$E_{MCA}$ is plotted as function of the Fe-O distance in Fig.~\ref{fig:2}(b) 
for the case of 6-ML, where the dotted line denotes $E_{MCA}$ of 6-ML Fe without MgO.
$E_{MCA}$ decreases as the Fe-O distance increases.
When the distance exceeds 3.0~\AA, MCA energy becomes equal to that of MgO-free Fe layers.
Definitely, this implies that hybridization affects $E_{MCA}$.
However, in forthcoming discussion,
we will show that the hybridization, though not completely ignorable,
is not the main contribution to PMCA as claimed.

The atomically decomposed $E_{MCA}$ is presented as function of number of Fe layer
in Fig.~\ref{fig:2}(c) and (d) for the clean-Fe and Fe/MgO, respectively.
In the clean-Fe, the surface contribution is largest, 
while those from the subsurface and the center layers are much smaller.
Contributions from the subsurface are even negative.
On the other hand, with MgO
the interface layers in contact with the MgO
contribute almost equally as the surface layer when the number of Fe layers exceeds three.
Contributions from layers other than interface and surface are negligibly small.
[See Supplementary Information].
If the $p$-$d$ hybridization is really the main source of the large PMCA observed in the Fe/MgO,
then why the surface Fe layer with no hybridization contributes almost equally as Fe(I)?
This question will be explored in forthcoming discussions.

\begin{figure}[t]
  \centering
  \includegraphics[width=\columnwidth]{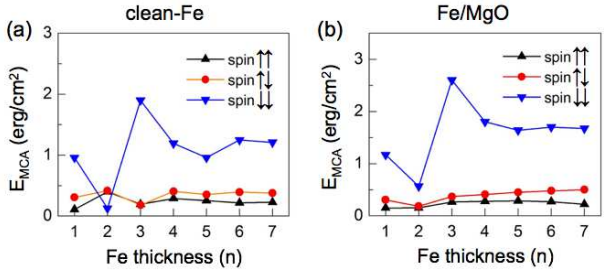} 
  \caption{(color online) Spin-channel decomposition of $E_{MCA}$ for (a) clean-Fe and (b) Fe/MgO,
  where red (blue) symbols denote $\uparrow\uparrow$ ($\downarrow\downarrow$) channel, 
  and black ones for $\uparrow\downarrow$ channel.}
  \label{fig:3}
\end{figure}

The spin-channel decomposition is presented in Fig.~\ref{fig:3}
for (a) the clean-Fe and (b) the Fe/MgO.
Again, it is plotted as function of number of Fe layers,
where the saturation behavior is evident.
Remarkably, in both cases$-$ with and without MgO,
the $\downarrow\downarrow$ channel dominates over the other channels
since the majority spin states are almost fully occupied
as in the case of Fe multilayers\cite{wang93:prb,wu99:jmmm}.

\begin{figure}[b]
  \centering
  \includegraphics[width=\columnwidth]{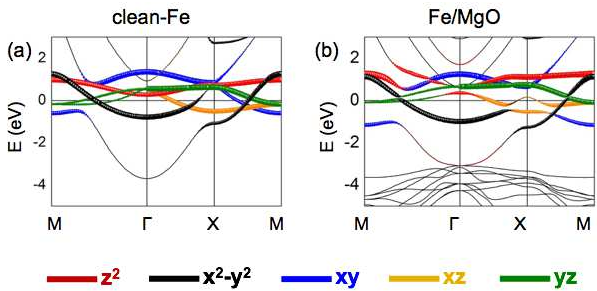}
\caption{(color online)
  Band structure of the free-standing Fe (1-ML Fe) and Fe/MgO.
  Only the minority spin bands are shown.
  (a) 1-ML Fe,and  (b) 1-ML Fe/MgO. 
  The orbital contribution of Fe $d$ state is emphasized in colors: 
  red ($d_{z^2})$, black ($d_{x^2-y^2}$), blue ($d_{xy}$), orange ($d_{xz}$), and green ($d_{yz}$), respectively.
}
  \label{fig:4}
\end{figure}

To elucidate the physics origin of PMCA,
band structures of Fe ML without and with MgO are shown in Fig.~\ref{fig:4}.
We first analyze the 1-ML case for simplicity,
to clarify the physics without loss of generality.
The overall bands look so similar in two cases.
The presence of MgO affects $d_{z^2}$ and $d_{xz,yz}$ orbitals.
Small shifts of $d_{xz,yz}$ are apparent due to MgO.
At $X$, $d_{xz}$ are split into two states, one 0.6 eV below $E_F$ and the other near $E_F$.
On the contrary, in-plane orbitals are less affected by the presence of MgO.

For the MCA analysis, we follow the recipe by Wang {\em et al.}\cite{wang93:prb},
which has proved to be successful
in various materials\cite{wang93:prb,wu99:jmmm,zhang10:prb82,odkhuu11:apl,odkhuu13:prb,odkhuu15:prb,hotta13:prl,hotta13:jkps}.
The increase of MCA by the presence of MgO can be viewed in two ways.
First,
positive contribution by $\langle xz|\ell_Z| yz\rangle$ remains the same 
in spite of litte reduced contribution by $\langle x^2-y^2|\ell_{Z}| xy \rangle$ around $\Gamma$.
Second, negative contributions
by $\langle z^2 |\ell_X| yz\rangle$
are reduced as the occupied $d_{yz}$ state becomes unoccupied about $\frac{1}{2}M$-$\Gamma$.
Third, negative contributions such as 
$\langle xy |\ell_X| xz,yz\rangle$ in $X$-$M$
are reduced owing to enlarged energy differences.
As such, the presence of MgO slightly affects some bands due to weak hybridization.
Nonetheless, the overall bands do not change much with a little increase in MCA.

\begin{figure}[t]
  \centering
  \includegraphics[width=\columnwidth]{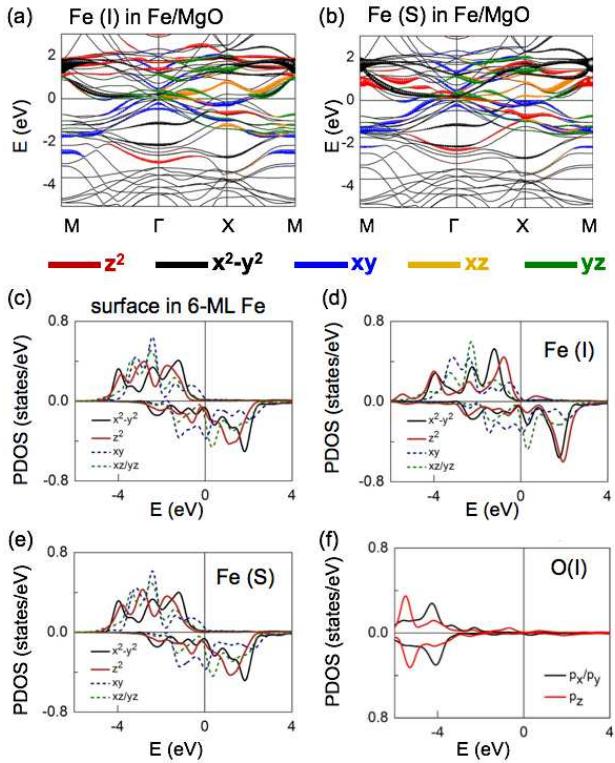}
  \caption{ (color online) Band structure of (a) Fe (I) and (b) Fe (S) in 6-ML Fe/MgO.
    Color representing $d$ orbitals are the same as Fig.~\ref{fig:4}.
    Density of states (PDOS) of Fe and Fe/MgO with 6-ML Fe.
    The surface Fe (c) without MgO, (d) Fe(I), (e) Fe(S)with MgO,
    and (f) O(I) $2p$ orbitals. 
    The $d$ orbital states are shown in different colors: 
    red ($d_{z^2})$, black ($d_{x^2-y^2}$), blue ($d_{xy}$), 
    and green ($d_{xz,yz}$), respectively.
    For the $2p$ orbitals: $p_x$, $p_y$, and $p_z$, respectively.}
  \label{fig:5}
\end{figure}

For completeness, the case of 6-ML Fe/MgO are analyzed.
Contributions from Fe(I) and Fe(S) to bands are shown
in Fig.~\ref{fig:5}(a) and (b), respectively.
Despite the complexity of bands,
it is noticeable that the energy levels of those orbitals involving
matrices $\langle x^2-y^2|\ell_Z|xy\rangle$ and $\langle xy|\ell_X|xz,yz\rangle$
do not differ very much in both Fe(I) and Fe(S). 
This implies that the hybridization is so weak that MCA from Fe(I) and Fe(S) are comparable.
The large PMCA in Fe/MgO is a result of interplay of $\ell_{Z}$ and $\ell_{X}$ matrices.

To get more insights on the role of Fe(S) and Fe(I),
DOS of the free-standing 6-ML Fe and the 6-ML Fe/MgO
are presented in Fig.~\ref{fig:5}(c-f),
where the left panel is for Fe(S) of the clean Fe and on MgO,
and the right panel for Fe(I) and O(I) in Fe/MgO.
DOS of Fe(S) is little affected by MgO.
The majority spin states of Fe(I) and Fe(S) are almost filled
with peaks from $d_{x^2-y^2}$ and $d_{z^2}$ 1$\sim$2 eV above $E_F$. 
In particular, Fe(I)
peaks from $d_{z^2}$ in the majority spin state are closer to $E_F$.
On the other hand, minority spin $d_{xy}$ and $d_{xz,yz}$ states close to $E_F$ form peaks.
The minority spin states of Fe(I) retain almost the same feature of the free-standing 6-ML Fe.
DOS of O(I) is also shown in Fig.~\ref{fig:5}(f),
where $p$ states are prominent around -6 $\sim$ -4 eV,
which is rather far away from Fe $d_{xz,yz}$ and $d_{z^2}$ states.
As clearly seen from the DOS plots, the hybridization is weak in Fe/MgO.

While MCA shows convergent behavior with the increase of the number of Fe layers,
the singular behavior of 2-ML is puzzling.
To tackle this issue, the unique structural feature of 2-ML should be emphasized.
Without MgO, both Fe layers are symmetrically equivalent.
On the other hand, with MgO, while one layer is the surface layer the other is the interface layer.
Moreover, the 2-ML exhibits rather unique electronic structure
[See Supplementary Information]:
$d_{x^2-y^2}$ state are prominent in DOS for both spin states just below $E_F$.
These states couple with the minority-spin $d_{xz,yz}$ states,
hence positive contribution is compensated by negative one.
Also, Fe 2$p_{1/2}$ core-levels are analyzed.
The core-level shifts reflect either charge transfer
or change of internal electric field\cite{weinert95:core}.
Regardless of the presence of MgO,
2$p_{1/2}$ energy of the center Fe layer is lowest,
while those of the interface and the surface are higher.
[See Supplementary Information for core-levels.]
In other cases, Fe layers are thick enough to screen out internal field by MgO,
which is not the case for 2-ML Fe, as manifested in Fe 2$p_{1/2}$ core levels.

So far in our calculations, the stoichiometry is assumed to be perfect.
On the other hand, in over- and under-oxidized Fe/MgO\cite{yang11:prb},
the Fe layer in contact with the over- {\em or} under-oxidized MgO
is expected to show ripples
Here, we argue that perfect epitaxy at the interface
is the key to achieve large PMCA in Fe/MgO.
Results clearly show that MCA contributions from Fe(S) and Fe(I) are similar in magnitudes
regardless of the hybridization.
Hence, the hybridization is weak as evidenced in DOS,
which cannot be the main driving force for large PMCA.
although its contribution to PMCA cannot completely neglected.
Before we conclude, we emphasize here that the physics we have demonstrated 
occurs in other lattice constants as well.
Even if we cannot exclude the strain effect in Fe/MgO,
we argue that the main driving force of PMCA is the perfect epitaxy
which is intrinsically achieved in {\em ab initio} calculations.

In summary,
we have shown that the $p$-$d$ hybridization, though not completely ignorable,
cannot be the main driving force for large PMCA,
which is different from commonly accepted.
The hybridization is overall weak, and indeed increases systematically PMCA with MgO.
However, MCA with and without hybridization are almost equal as well manifested in Fe(S) and Fe(I).
From the detailed analysis, contributions from the surface and the interface Fe are dominant.
As in the Fe multilayers, since $d$ states of the majority spin bands are completely filled,
it is the $\downarrow\downarrow$ channel that plays the most dominant role in PMCA.
Furthermore, while the presence of MgO retains positive contribution by $\langle xz|\ell_Z|yz\rangle$,
negative contribution $\langle z^2|\ell_X|yz\rangle$ and $\langle xy|\ell_X|xz,yz\rangle$ are slightly reduced.
More importantly, perfect epitaxy,
which is an indicator of high level of crystal growth,
is the key factor to realize large PMCA in Fe/MgO
from the fact that Fe layer with and without hybridization 
contribute almost equally to PMCA.

This work was supported by the Priority Research Centers
Program through NRF funded by theMinistry of Education of
Korea (2009-0093818) and the Basic Science Research Program
through NRF funded by the Ministry of Science, ICT
and Future Planning (2015R1A2A2A01003621).

%

\end{document}